\documentclass[iop,apjl,tighten]{emulateapj}
\usepackage{amssymb}
\usepackage{epsfig}
\usepackage{graphics}
\usepackage{natbib}

\usepackage{cprotect}

\newcommand{\rmd}{{d}}   
\newcommand{\me}{m_{\rm e}}   

\newcommand{\hii}{\mbox{H\,{\sc ii}}}
\newcommand{\heii}{\mbox{He\,{\sc ii}}}

\slugcomment{ApJ, in press}

\begin{document}

\title{The mystery of spectral breaks: 
Lyman continuum absorption  by photon-photon pair production 
in the \textit{Fermi} GeV  spectra of bright blazars} 

\shorttitle{GeV breaks in \textit{Fermi} blazar spectra}
\shortauthors{Stern \& Poutanen}

\author{Boris E. Stern\altaffilmark{1,2} and Juri Poutanen\altaffilmark{3,4} } 

\affil{
$^1$Institute for Nuclear Research, Russian Academy of Sciences, Prospekt 60-letiya Oktyabrya 7a, Moscow 117312, Russia; stern.boris@gmail.com \\
$^2$Astro Space Center, Lebedev Physical Institute, Russian Academy of Sciences, Profsoyuznaya 84/32,  Moscow 117997, Russia \\
$^3$Tuorla Observatory, University of Turku, V\"ais\"al\"antie 20,   FI-21500 Piikki\"o, Finland;  juri.poutanen@utu.fi \\ 
$^4$Astronomy Division, Department of Physics, PO Box 3000, FI-90014 University of Oulu, Finland }  

\date{Received 2013 December 19; accepted 2014 August 3} 

\begin{abstract} 
\noindent 
We reanalyze \textit{Fermi}/LAT gamma-ray spectra of bright blazars with a higher 
photon statistics than in previous works and with new Pass 7 data representation. 
In the spectra of the brightest blazar 3C~454.3 and possibly of 4C~+21.35  
we detect breaks at $\sim$5 GeV (in the rest frame) 
associated with the photon-photon pair production absorption by  \heii\ Lyman continuum (LyC). 
We also detect confident breaks at $\sim$20 GeV associated with hydrogen LyC both in the individual spectra 
and in the stacked redshift-corrected spectrum of several bright blazars. 
The detected breaks in the stacked spectra univocally prove that they are associated with atomic ultraviolet  
emission features of the quasar broad-line region (BLR). 
The dominance of the absorption by hydrogen Ly complex over \heii,   
rather small detected optical depth, and the break energy consistent with the head-on collisions with LyC photons
imply that the gamma-ray emission site is located within the BLR, but 
most of the BLR emission comes from a flat disk-like structure producing little opacity. 
Alternatively, the  LyC emission region size might be larger than the BLR size measured from reverberation mapping, 
and/or the $\gamma$-ray emitting region is extended. 
These solutions would  resolve a long-standing issue how the multi-hundred GeV photons can escape from the emission 
zone without being absorbed by softer photons.  
\end{abstract}



\section{Introduction} 

The spectra of bright blazars obtained by \textit{Fermi Gamma-ray Space Telescope} (\textit{Fermi}) Large Area Telescope (LAT) 
showed clear deviations from a power-law shape \citep{Abdo09_3C454.3,Abdo10_blazars}. 
There spectra could not be described by smooth functions such as exponentially cutoff power-law or 
a log-parabola (log-normal distribution), but were found to be better described by a broken power-law. 
The derived break energies lying in the 1--10 GeV energy range \citep{Abdo10_blazars,PS10,Harris12} 
were rather stable \citep{Fermi10_3C454,Fermi11_3C454,SP11}.  
Those breaks seemed puzzling: the hypothesis that the break is caused by photon-photon 
annihilation through $e^\pm$ pair production was considered and rejected by \citet{Abdo10_blazars}, 
who  argued that 
``to produce a break in the 1--10 GeV, the photon field should have an energy peaking in
the 0.05--0.5 keV range, which excludes the broad-line region peaking in the UV.'' 
The conclusion that such large energies of the target photons  were required was based on an erroneous assumption 
that the break energy correspond to the maximum cross-section for pair production. 

\citet{PS10} suggested that the breaks should actually appear at the energies close to 
the threshold for a corresponding reaction, where the opacity has a sharp rise.  
The observed breaks at a few GeV correspond then well (correcting for the redshift) to  
the Lyman recombination continuum (LyC) and Ly$\alpha$ emission of ionized  He. 
They also showed that the inner part of the BLR can provide sufficient flux of \heii\  Lyman lines and LyC 
to provide enough opacity for GeV photons and to produce a spectral break.
\citet{PS10} further demonstrated that the data for a number of bright blazars are well 
described by a power-law spectrum modified by the absorption within the BLR. 
The fits with this model were acceptable and the reduction in $\chi^2$ (compared to a simple power-law model) was very significant. 
Similar $\chi^2$ could also be achieved with the broken power-law model which, however, does not have any physical basis. 

A high significance of the spectral breaks partially results from a null hypothesis for the underlying spectrum, 
which is assumed to be a power-law. 
However,  a typical blazar has a curved spectrum extending over many orders in energy  and peaking in the MeV--GeV range.  
This implies that  the spectrum in the \textit{Fermi} energy band should be slightly concave   
and the power-law null hypothesis gives an overestimated significance of the break. 
As a more realistic null hypothesis one can take a lognormal distribution (log-parabola), 
which is the simplest way to introduce a curvature in logarithmic coordinates. 

\citet{SP11} studied in details the spectrum of the exceptionally bright flat-spectrum radio quasar (FSRQ) 3C~454.3 
and indeed found that the spectrum below the absorption break significantly differs from the power-law 
and can  be  well described by a log-parabola. 
With this null hypothesis the statistical significance of any break is lower than with a power-law hypothesis. 
However, the absorption break in the time-integrated spectrum of 3C~454.3 was still highly significant and 
its energy coincided with the predicted one from \heii\  LyC absorption. 

While the main attention of the cited papers has been payed to the GeV breaks, 
\citet{PS10} and \citet{SP11} also revealed hydrogen LyC breaks at $\sim$20 GeV,  
which were less significant and less impressive because of a lower photon statistics in that energy range. 
Actually, it is obvious that in most cases 
H Ly radiation should produce a stronger absorption feature that \heii\  Ly emission. 
That is why it is important to revisit the spectral analysis of bright blazars with a higher statistics that was accumulated since previous works. 
Another, more important reason to revisit previous results is a slight but significant difference in the blazar spectra 
obtained with the new Pass 7 version of the \textit{Fermi}/LAT data and detector response and those obtained with the older Pass 6 version. 
The new spectra look smoother and this is a reason to suspect that the sharp breaks at a few GeV were the 
artifacts of the Pass 6 response function -- partially or completely.

\begin{figure}
 \plotone{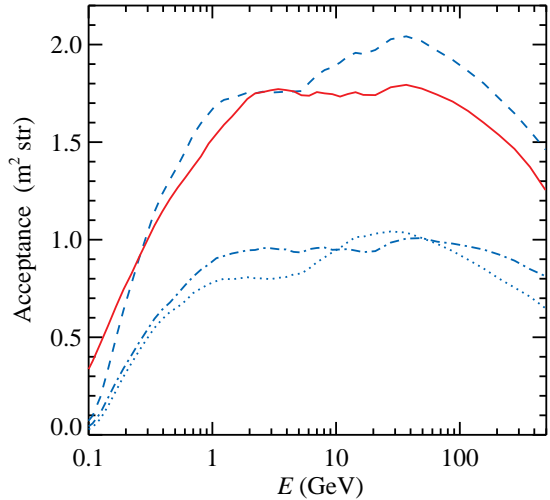}
\cprotect\caption{Acceptance (i.e. effective area integrated over the field of view) of \textit{Fermi}/LAT.
The solid red curve gives the acceptance for Pass 7 version {\verb|P7_CLEAN_V6|}.
The blue  curves are for Pass 6  version \verb|P6_V3_DIFFUSE|:
the dashed upper curve is the total acceptance for two detectors, the dotted 
curve is for the front detector and the dot-dashed curve is  for the back detector. 
The acceptance of the  front detector has a hump starting at $\sim$3 GeV and peaking at 30 GeV,
which enhances the \heii\  absorption feature or even mimics it.}
\label{fig:acceptance}
\end{figure}

\section{Data and their analysis} 
\label{sec:data}

We use \textit{Fermi}/LAT photon data for 1740 days (from 2008 August 6 till 2013 May 12) using the new Pass 7 event classification, 
selecting photons of  clean class, and imposing the cut on the zenith angle at $\theta < 105 \degr$.    
We used \verb|P7_CLEAN_V6| response function.  
The diffuse background was calculated using the background model 
elaborated by the \textit{Fermi} team.

\begin{figure}
 \plotone{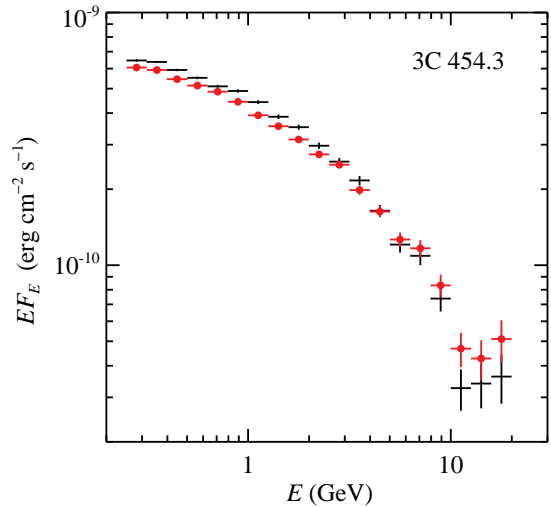}
\cprotect\caption{Central part of spectra of 3C454.3 using 
response functions for Pass 6 version \verb|P6_V3_DIFFUSE| (black crosses) 
and Pass 7  version {\verb|P7_CLEAN_V6|} (red circles and crosses). 
}
\label{fig:3C454}
\end{figure}

It should be noted that the Pass 7 response function significantly differs from the Pass 6 one in at least two aspects:
\begin{enumerate}
\item it has a wider point spread function,
\item it has a different effective area function (see Fig.~\ref{fig:acceptance}). 
\end{enumerate}
The second fact is of crucial importance for the analysis of the GeV energy breaks because  
the Pass 6 response has a hump starting at 4--5 GeV, which introduces a break in the 
photon spectrum at energies close to the \heii\  LyC absorption 
(see Fig.~\ref{fig:3C454} for comparison of the spectra of 3C~454.3 obtained with Passes 6 and 7). 
The break is not very strong: it changes the index of the power-law  by $\sim$0.1. 
If one then fits  the resulting spectrum with a broken power law, one obtains a significant break. 
We believe that the break in the Pass 6 effective area is not real, because there is no clear reason for existence of such a feature at this energy. 
Moreover, it looks strange that the hump appears only in the effective area of the front detector. 
The Pass 7 version of the effective area does not have such a feature and looks more natural.

The energy-dependent exposure function was calculated using the spacecraft pointing history: 
\begin{equation}
\mbox{Exposure}(E,\Delta T) = \int_{\Delta T} S(E,\theta(t)) \ \rmd t,
\end{equation}  
where $\Delta T$ is the time interval of interest, 
$\theta$ is the angle between the detector axis and the direction to the object and 
$S (E,\theta)$ is the detector effective area at energy $E$. 
We accumulated counts in the circle centered at the source location with the energy-dependent radius 
$r(E)=\min\{r_{90}, 4\degr\}$, where $r_{90}$ is the radius of 90\% event containment, which was calculated 
with  Monte-Carlo integration of the point-spread function assuming the isotropic distribution of the exposure angle. 
Then number of counts in each energy bin was corrected to the containment factor for $r(E)$. 

In order to reveal the absorption features in the blazar emission, 
we analyzed the spectra of individual bright FSRQs as well as the stacked spectra of various samples.   
We have selected a sample of 15 brightest objects from the 2nd Fermi catalog \citep{Nolan12} 
using the  following criteria:
\begin{enumerate}
\item  the total number of counts above 1 GeV after background subtraction is above 1800,
\item  the signal/background ratio exceeds 2,
\item  known redshift (according to the 2nd Fermi catalog),
\item  classification as a FSRQ, or as a low-synchrotron peak  BL~Lac if its redshift exceeds 0.5 (which would mean that 
the latter is probably a misclassified FSRQ), 
\item  there is no strong source confusion.
\end{enumerate}

The brightest blazar 3C~454.3 was excluded from the stacking analysis and studied only individually because of its exceptional brightness, 
which is comparable to the total signal from other selected objects. 
The remaining 14 FSRQs bright above 1 GeV constitute Group 1, where we hoped to reveal the absorption break at $\sim$5 GeV 
associated with \heii\  LyC emission \citep{PS10}. 

We also selected Group 2 of eight blazars with the highest number of counts above 5 GeV (more than 200 after background subtraction).
This sample was more promising to study H LyC absorption break at $\sim$20 GeV.  
All such blazars also belong to Group 1 despite the selection was independent.

\begin{deluxetable}{llcr}
\tabletypesize{\scriptsize}
\tablecaption{The brightest GeV blazars \label{tab:groups}}
\tablewidth{0pt}
\tablehead{
\colhead{Object} & \colhead{Group\tablenotemark{a}} & \colhead{Redshift}  & \colhead{Dates\tablenotemark{b}}   }
\startdata      
3C~454.3      & &  0.859      & 0--1740  \\           
4C +55.17    &  1, 2   &0.896  & 0--1740  \\
PKS 0537$-$441& 1, 2 &  0.892 & 0--1740\\
PKS 2326$-$502  & 1 &  0.518 & 600--1740  \\
4C +21.35 (PKS 1222+21) & 1, 2 & 0.433& 350--1100 \\
PKS B1424$-$418 &1, 2 & 1.522& 1200--1740  \\
PKS 0426$-$380 &1, 2  & 1.111& 0--1740 \\
PKS 0454$-$234 & 1, 2 & 1.003& 0--1740 \\
PKS 0727$-$11 & 1 & 1.591& 0--\ 800 \\
PKS 1510$-$08 & 1, 2 & 0.360& 0--1740\\
3C 279 &1  &  0.536& 0--1300 \\
PKS 1502+106 & 1, 2 & 1.893& 0--\ 500 \\
B2 1520+31&1  & 1.484& 0--1400 \\
PKS 0235+164& 1 & 0.940 & 0--\ 400 \\
4C +38.41 & 1 & 1.813 & 0--1740 \\
 \multicolumn{4}{c}{BL Lacs}  \\
Mrk~421& 3 & 0.030 & 0--1740 \\
3C~66A & 3 & 0.444 & 0--1740 \\
S5~0716+714 & 3 & 0.310 & 0--1740 \\
PKS~2155$-$304  & 3 & 0.117 & 0--1740 \\
\enddata
\tablenotetext{a}{Group memberships. } 
\tablenotetext{b}{The start and the end of the observation measured from MJD 54684 (2008 August 6). }
\end{deluxetable}


To optimize the signal-to-noise ratio, we have selected for each blazar a time interval, 
when its flux substantially exceeded the background. Limits of these time intervals are given in Table~\ref{tab:groups}.
In order to prepare the stacked spectra, we first derived individual spectra using the energy bins with the width of 0.1 in decimal logarithm. 
The bin edges were adjusted in such a way that they are the same in the object rest frame. 
Each spectrum was blue-shifted by a factor $(1+z)$. 
We summed up the obtained spectra with their absolute normalization, so that the spectra of the brighter objects have a larger contribution. 
Such approach optimizes relative statistical errors.
The blazar spectra were modeled with a lognormal function with the superimposed absorption by the BLR emission (see Sect.~\ref{sec:opacity}).  
We use the opacity computed for different ionization  parameters  $\xi$ of the BLR as described in \citet{PS10}. 
We also checked a simpler monochromatic absorber model (H and \heii\ LyC),  
which is less realistic but helps to estimate separate contributions of H and He emission.

We have also constructed two comparison spectra, where breaks are not expected, 
to make sure that the detection of the breaks is not an artifact of the detector response. 
The first one is the  stacked redshift-corrected spectrum of four bright BL Lacs (Group 3 in Table~\ref{tab:groups}). 
The second one is the  ``empty'' sky spectrum. 
The sky was sliced into 72\,000 bins of equal area ($1.74\times10^{-4}$~str) 
and the photons were collected from the bins where number of photons above 100~MeV is less that 300, 
these bins constitute  about 0.226  fraction of the sky. 
The main contribution to this spectrum is from the high-energy protons producing pions and a smaller contribution of unresolved BL Lacs.

Statistical errors were treated as Gaussian, except in a few bins at higher energies, 
where the number of photons is low, we use Poisson likelihood adding $-2\ln P(n,\mu)$ to $\chi^2$ 
(here $n$ is the number of counts in the bin and $\mu$ is the prediction of the model). 
The number of such bins is small and the  meaning of $\chi^2$ is not significantly affected. 
For the minimization we use the standard code {\sc minuit} from the CERN library.

\section{Gamma-ray opacity} 
\label{sec:opacity}

Gamma-ray photons emitted presumably by the relativistic jet emanating from the black hole 
are strongly beamed. They propagate through the radiation field made by the accretion disk, the BLR and the dusty torus 
are potentially can be absorbed by  photon-photon pair production. 
The disk radiation moves nearly parallel to the photon beam and therefore does not interact efficiently. 
The infrared photons from the dust absorb radiation mostly in the TeV range. 
Thus, the most important source of opacity is the optical/UV (nearly) isotropic radiation from the BLR.  

The opacity depends on the BLR spectrum, which is computed using spectral synthesis code {\sc xstar} \citep[version 2.2,][]{KB01} 
as described by \citet{PS10}. 
We repeat here the basic assumption for completeness. 
The ionizing  spectrum of a quasar is taken as a sum of the standard multi-color accretion disk  
plus a power law of total luminosity 10\% extending to 100 keV \citep{Laor97}. 
The BLR clouds are assumed to be simple slabs of constant gas density and a clear view to the ionizing source. 
The hydrogen column density was fixed at $N_{\rm H}=10^{23}$ cm$^{-2}$ and 
the BLR spectra were computed for different ionization parameters  $\xi=L/(r^2 n_{\rm H})$  varying from 10$^{0.5}$ to 10$^{2.5}$.   
If one assumes  a  dependence of the cloud density on distance  from the black hole $n_{\rm H}= 10^{10}r_{18}^{-1}$, 
then $\xi=10\ L_{47} r_{18}^{-1}$ and our ionization range would correspond  to the distance interval between 
1 and 0.01 pc to the ionizing source for a quasar luminosity $L=10^{47}$~erg~s$^{-1}$.\footnote{We defined $Q=10^xQ_x$ in cgs units.}  
The scaling and the estimated distances are very approximate and are model-dependent. 

If the BLR were to emit  only one line at energy $E_0$, 
the optical depth  for a  $\gamma$-ray photon of energy $E$ through the region of size $R$ filled with isotropic soft 
photon field of column density $N_{\rm ph}$ would be  
$\tau_{\gamma\gamma}(E,E_0) = {\sigma_{\gamma\gamma} (s)}\  \tau_{\rm T} /{\sigma_{\rm T}} = N_{\rm ph} \sigma_{\gamma\gamma} (s) $, where 
\begin{equation} \label{eq:taut}
\tau_{\rm T} =  N_{\rm ph}  \sigma_{\rm T} =  \frac{L_{\rm line} \sigma_{\rm T}}{4\pi R c E_0} = 110 \frac{L_{{\rm line}, 45}}{R_{18}} \frac{10\ {\rm eV}}{E_0} ,
 \end{equation} 
$\sigma_{\rm T}$ is the Thomson cross-section,  $L_{\rm line}$ is the line luminosity, $s=EE_0/(\me c^2)^2$ and 
$\sigma_{\gamma\gamma}$ is the angle-averaged cross-section for photon-photon pair production \citep[see, e.g.][]{GS67,ZDZ88}. 
Note, that $\sigma_{\gamma\gamma}$ has  threshold $s=1$ 
(i.e. at $E_{\rm thr}=19.2\ \mbox{GeV} /(E_0/13.6\ \mbox{eV})$) 
and has a peak of about $\sigma_{\rm T}/5$ at $E\approx 3.5\ E_{\rm thr}$.

The BLR spectrum, of course, contains many lines and recombination continua. 
One can introduce  the   cross-section  weighted with the photon distribution: 
\begin{equation} \label{eq:mean sigma}
\overline{\sigma}_{\gamma\gamma} (E)  = \frac{1}{N_{\rm ph}} \int_{s>1} \sigma_{\gamma\gamma}  (s) N_{\rm ph}(E_0) {\rm d} E_0,   
\end{equation}
where $N_{\rm ph}=\int N_{\rm ph}(E_0) \rmd E_0$. 
The spectrum transmitted through the BLR is attenuated as $\propto \exp(-\tau_{\gamma\gamma}(E))$, 
where $\tau_{\gamma\gamma}(E)=\tau_{\rm T}\overline{\sigma}_{\gamma\gamma} (E)  /{\sigma_{\rm T}} $,
and $\tau_{\rm T}$ is computed using Equation~(\ref{eq:taut}) replacing $L_{\rm line}$ by $L_{\rm BLR}$  and 
 $E_0$ by the mean photon energy of the BLR: 
\begin{equation}
\overline{E}   = \frac{1}{N_{\rm ph}} \int  E_0 N_{\rm ph}(E_0) {\rm d} E_0.
\end{equation}

\begin{figure}
\plotone{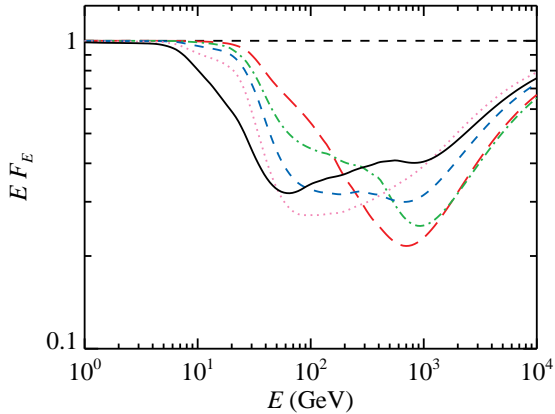}
\caption{Examples of the photon spectrum transmitted through the BLR of various ionizations and 
optical depths. The incident spectrum  (dashed black line) is taken as a power-law of photon index $\Gamma=2$. 
The total photon column density corresponds to $\tau_{\rm T}=10$ in all cases. 
Transmission function $\exp(-\tau_{\gamma\gamma}(E))$ for different $\log\xi$ is shown by different lines: 
0.5 (red long-dashed), 1.0 (green dot-dashed), 1.5 (blue short-dashed), 2.0 (pink dotted), 2.5 (black solid). 
}
\label{fig:blr_abs}
\end{figure}

As an illustration we present the results of absorption of a power-law spectrum 
by the BLR of different ionizations in Fig.~\ref{fig:blr_abs} fixing the total BLR photon column density 
at $N_{\rm ph}=1.5\times 10^{25}$ cm$^{-2}$, which corresponds to $\tau_{\rm T}=10$.  
For the considered $\tau_{\rm T}$, 
the flux drops at most by a factor of 3--4.5 depending on $\xi$, corresponding to 
the maximum optical depth of about 1.1--1.5.  
Note that the transmitted spectrum in the range from 30~GeV to 1 TeV 
has nearly the same slope as the intrinsic one    at larger $\xi$, because the opacity is nearly constant  in this range. 
The opacity drops at energies above  1 TeV and the spectrum recovers. 
We see that the \heii\ LyC breaks at 5 GeV are more pronounced at high ionizations $\log\xi>1.5$, 
while the H LyC breaks are seen at any $\log\xi$. 
This allowed \citet{PS10} to introduce a simpler double-absorber model 
for $\gamma$-ray opacity, where the BLR spectrum is replaced by the strongest 
emission features of H and \heii\ LyC. For low ionization, one can even consider only 
a single absorber due to the H LyC.

\newpage
\section{Results} 
\label{sec:results}

\subsection{Detection of GeV breaks} 

\begin{figure}
\plotone{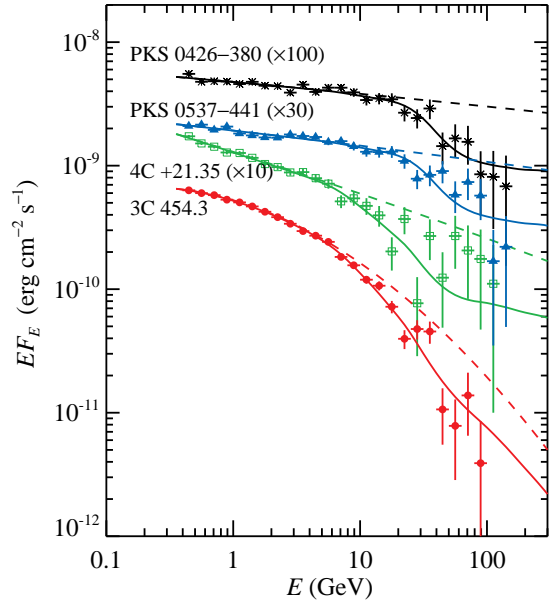}
\caption{The redshift-corrected \textit{Fermi}/LAT spectra of individual bright blazars    
and their best-fit model of the log-normal distribution with absorption by the BLR (with $\log\xi=1.5$).
The dashed lines show the  same log-normal distributions without absorption.
}
\label{fig:spe_individ}
\end{figure}

\begin{figure}
 \plotone{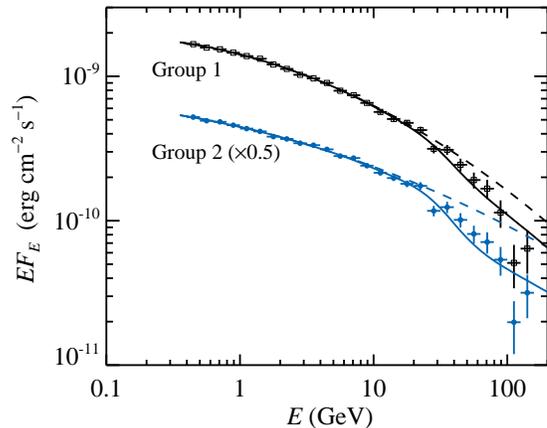}
\caption{Same as Fig.~\ref{fig:spe_individ}, but for 
the stacked rest-frame  spectra for the two samples of blazars from Table~\ref{tab:groups} 
for 1740 days of \textit{Fermi} observations. 
}
\label{fig:spe_stacked}
\end{figure}

\begin{deluxetable*}{llccccccccc}
\tabletypesize{\scriptsize}
\tablecaption{Spectral Properties of Blazars \label{tab:fits}}
\tablewidth{0pt}
\tablehead{                                    
\colhead{Object} & \colhead{lognorm}  
&  \multicolumn{2}{c}{lognorm+H LyC\tablenotemark{a}} 
&  \multicolumn{2}{c}{lognorm+H\&\heii\ LyC\tablenotemark{b}}  
&  \multicolumn{2}{c}{lognorm+$\log \xi$=1.5\tablenotemark{c}}   
& \multicolumn{2}{c}{lognorm+$\log \xi$=2.5\tablenotemark{d}}  &  \\
                                          & \colhead{$\chi^2$/dof\tablenotemark{e}} 
& \colhead{$\chi^2$/dof\tablenotemark{e}}   & \colhead{$\tau_{\rm H}$\tablenotemark{f}} 
& \colhead{$\chi^2$/dof\tablenotemark{e}}   & \colhead{$\tau_{\rm He \mbox{\sc ii}}$\tablenotemark{g}}  
& \colhead{$\chi^2$/dof\tablenotemark{e}}   & \colhead{$\tau_{\rm T}$}  
& \colhead{$\chi^2$/dof\tablenotemark{e}}   & \colhead{$\tau_{\rm T}$}  
&  \colhead{Significance\tablenotemark{h}} }     
\startdata                        
3C~454.3                &    55.0/21  &      38.3/20  & 4.4$\pm$1.0   &       28.1/19 & 0.94$\pm$0.3 &      29.6/20   & 14.0$\pm$4.2 &     25.8/19  & 8.8 $\pm$1.7 & 5.5$\sigma$\\
PKS~B1424$-$418 &   23.0/23    &     19.0/22  & 2.0$\pm$1.0 &      19.0/19 & $<$0.3    &         18.0/20   & 6.1$\pm$2.9  &    23.8/20  & $<$4.2 & \nodata  \\
PKS 0426$-$380    &  42.3/23   &     27.5/22  & 3.8$\pm$0.7 &      27.5/21  & $<$0.4   &        22.9/22   & 9.6$\pm$1.7  &    36.4/22  & 6.3$\pm$ 2.6& 4.5$\sigma$\\
PKS 1502+106       &  30.5/21   &    22.5/20  & 3.2$\pm$1.2  &      22.5/19 & $<$0.17   &        21.1/20   & 9.0$\pm$3.3  &     30.2/20  & 1.5$\pm$1.3 & 3$\sigma$\\
PKS 0537$-$441   &  46.0/23   &  34.1/22  & 3.2$^{+1.5}_{-1.0}$ &  34.1/21 & $<$1.4     &       29.3/22    & 9.1$\pm$1.5 &      40.6/22  & 5.5$\pm$2.6 & 4$\sigma$\\
PKS 0454$-$234   &  35.7/23   &   28.7/22  & 3.7$\pm$1.5    &      27.5/21 & $<$0.4    &        28.2/22    & 10.6$\pm$4.5 &     34.1/22   & 4.2$\pm$3.4 & 2.5$\sigma$ \\
4C +21.35             &  35.7/22   &     33.1/21  & 3.3$\pm$2.3  & 25.5/20  & 1.8$_{-0.5}^{+1.0}$ &  33.4/21   & 25$_{-21}^{+14}$ &  23.8/21  & 11.2$\pm$2.5& 3.5$\sigma$ \\  
PKS 1510$-$08    &   20.0/21 &    20.0/20   &$<$0.5         &    20.0/19  & $<$0.5       &      20.0/20    & $<$1.4       &      20.0/20  & $<$0.9 & \nodata   \\
4C +55.17            &    66.0/21    &    57.9/20     & 3.1$\pm$1.2   &      57.0/19 & $<$0.8       &      57.0/20   & 7.9$\pm$2.7  &      58.0/20  & 5.6$\pm$1.2 & \nodata \\
Group 1               &   44.0/23   &      36.8/22 & 1.0$\pm$0.35&     36.4/21 & $<$0.05     &        30.2/22 & 3.4$\pm$1.0 &      44.0/22  & $<$1.2 &  3.5$\sigma$ \\
Group 2               &   65.6/23   &    42.6/22  & 2.0$\pm$0.4 &      42.6/21 & $<$0.1     &         31.6/22  & 6.2$\pm$1.1  &    52.9/22 & 2.9$\pm$1.2 & 6$\sigma$   \\
Group 3 (BL Lacs) & 35.4/22 &  33.7/21 & 0.37$\pm$0.3 & 33.7/20  & $<$0.1 & \dots & \dots & \dots & \dots & \dots 
\enddata
\tablenotetext{a}{The lognormal distribution with a single H LyC absorber. } 
\tablenotetext{b}{The lognormal distribution with a double absorber by H and \heii\ LyC. } 
\tablenotetext{c}{The lognormal distribution with absorption provided by the BLR spectrum with ionization parameter $\log \xi=1.5$ 
\citep[see Sect.~\ref{sec:opacity} and][]{PS10}. } 
\tablenotetext{d}{Same as case c, but for $\log \xi=2.5$. } 
\tablenotetext{e}{The number of degrees of freedom (dof) differs because the spectra are cut at the first  bin with negative flux.} 
\tablenotetext{f}{Optical depth $\tau_{\rm T}$ due to H LyC only. } 
\tablenotetext{g}{Optical depth $\tau_{\rm T}$ due to \heii\ LyC only. } 
\tablenotetext{h}{The significance of $\chi^2$ reduction of the best-fit model with respect to the  fits with lognormal function. } 
\end{deluxetable*}

The results of the spectral fits for 3C~454.3, all objects of Group 2 and for the stacked spectra are presented in Table~\ref{tab:fits} and 
some of them are shown in Figs. \ref{fig:spe_individ} and \ref{fig:spe_stacked}. 
The best-fit model for all objects except 3C~454.3 and 4C~+21.35 is that of the BLR emission with lower ionization degree $\log \xi$ = 1.5. 
In this ionization state, the contribution by  \heii\  absorption is small and one can see from Table~\ref{tab:fits} 
that the double absorber model H+\heii~LyC does not improve the fits with respect to the single H~LyC absorber. 
This means that in most spectra there is no sign of \heii~LyC absorption. 
The exceptions are 3C~454.3 and 4C +21.35 (PKS~1222+21) where the presence of  \heii\  absorption is detected at $\sim 3\sigma$ level. 
The best-fit model for absorber in these sources is BLR emission with $\log \xi = 2.5$.  
In the stacked spectra of both groups, the situation is similar: the addition of  \heii\  does not change $\chi^2$ significantly.

The typical optical depth, $\tau_{\rm T}$, for the best-fit BLR emission model (mostly with $\log\xi=1.5$) was measured to between 4 and 20. 
This corresponds to the maximum optical depth of about 0.4--2.2 (see blue dashed line in Fig.~\ref{fig:blr_abs}) and 
the flux reduction at $\sim$100 GeV by a factor of 1.5--9.  
To estimate the absorption optical depth that is contributed by H and \heii\  emission only, 
one can consider corresponding optical depths from single- or double-absorber models 
(see Table~\ref{tab:fits}, columns 4 and~6).

Spectra of the six from the nine brightest (above 5 GeV) blazars demonstrate clear absorption breaks dominated by the H LyC absorption. 
The significance of these breaks ranges from 2.5$\sigma$ to 5.5 $\sigma$. 
The typical optical depth by H LyC only is  $\tau_{\rm H}\sim$2--4, 
which can be converted directly to the column density of LyC (plus Ly$\alpha$) photons 
on the line of sight to the $\gamma$-ray emitting region $N_{\rm ph, H LyC}=\tau_{\rm H}/\sigma_{\rm T}\approx (3-6) \times 10^{24}$.  
The stacked spectrum of Group 2 (which does not include 3C~454.3) has 6$\sigma$ significance of the break. 
This is the most confident and the most conservative demonstration that the H LyC absorption in bright blazars is a very typical phenomenon. 
Previously \citet{PS10}, \citet{SP11}, and recently  \citet{Tanaka13} revealed this absorption in individual sources. 
However, in first and in the last papers a less conservative assumption of a power law null hypothesis was used. 

As for the GeV breaks associated with \heii\ absorption (now with a more accurate Pass 7 effective area), 
they get a status of a rare phenomenon. 
There are indications of such breaks in two objects: 3C~454.3 (3$\sigma$) and 
4C~+21.35 (almost $3\sigma$), see Fig.~\ref{fig:spe_individ}.

\subsection{Significance of the breaks} 

Detection of the breaks due to H LyC absorption has high significance. 
The amplitude of the spectral deviation from the null hypothesis is factor of 2 in the case of the stacked spectrum 
(Group 2) and factor of 3 in the case of PKS 0426$-$380.  
This is much above possible systematic errors like uncertainties in the response function.

The breaks due to \heii~LyC are much weaker than was claimed before \citep{Fermi10_3C454,Fermi11_3C454,SP11} 
mostly because of the difference between Pass 6 and Pass 7 detector response functions. 
A sharp rise in the Pass 6 effective area (see Fig.~\ref{fig:acceptance}) made the spectral break observed at 2--3 GeV 
sharper and more significant. 
However, we believe that the breaks observed in two objects, 3C~454.3 and  4C~+21.35, are real. 
First, now there is no any feature in the LAT response function at the corresponding energy. 
Second, deviation of the spectra in the rest-frame range 5--20 GeV  
from the extrapolation of the lognormal function fitted to the data below 5~GeV reach at least 50\% 
(see Fig.~\ref{fig:spe_individ}), much above the uncertainty in the LAT response. 
It should be noted that the 3$\sigma$ significance level is quite serious in this case, because the detection of the effect 
does not include ``hidden trials''  like thresholds adjustments and sample manipulation. 
This significance is also very conservative as it was measured for the lognormal null hypothesis, not a power-law. 
For 3C~454.3, we do not expect systematical errors associated with the background subtraction 
or source confusion due to exceptionally high $\gamma$-ray brightness. 
4C~+21.35 is weaker and there could be some systematics, e.g., an underestimated soft background: 
the soft part of the spectrum could be lower and the spectrum could then be fitted with a narrower lognormal distribution 
without absorption. Therefore, the \heii\ absorption break in 4C~+21.35 needs further studies.

\begin{figure}
 \plotone{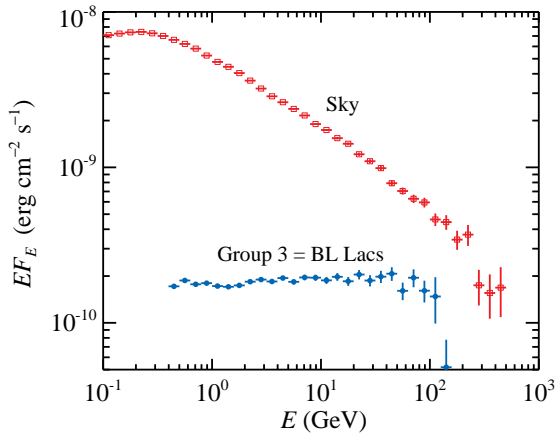}
\caption{Spectral energy distribution of the ``empty'' sky is shown by red open squares. 
The stacked spectrum of four brightest BL Lacs (Group 3: Mrk~421, 3C~66A, S5~0716+714 and PKS~2155$-$304)
is shown by blue circles. 
Neither of the spectra show any signs of absorption by H or He LyC in the range 2--20 GeV.
}
\label{fig:sky_bllac}
\end{figure}

In order to prove the presence of the GeV breaks in FSRQ, we checked whether 
the breaks appear also in the comparison spectra (see Fig.~\ref{fig:sky_bllac}), where they are not expected.  
We see that the stacked BL Lac spectrum (Group 3) is well described by a power-law 
(with some deviations because of imperfectness of the response function) without any breaks 
(except possibly at 100~GeV). 
The upper limits on the opacity due to H or He LyC are significantly 
below the detected opacity in the bright blazars (see Table~\ref{tab:fits}). 
The ``empty'' sky spectrum also does not show any signs of the breaks in the 2--20 GeV range. 
Thus, inaccuracies in the  detector response are unlikely to affect our conclusions.

We also made a simple test whether a different null hypothesis for the underlying spectrum 
would remove the necessity of the breaks. 
Instead of the log-normal function, we assumed, following the referee's suggestion, a biquadratic function 
$EF(E)\propto \exp[- A \ln^2(E/E_{\rm peak}) - B \ln^4(E/E_{\rm peak}) ]$, 
which has the same number of free parameters as our log-normal distribution 
with BLR absorption.  
This ad hoc function gives a slightly better fit with $\chi^2/{\rm dof}=20/19$ for 3C~454.3, but 
a much worse fit with $\chi^2/{\rm dof}=41/22$ for the Group 2 spectrum (see Table~\ref{tab:fits}). 
Interestingly, a fit for 3C~454.3 gives $A\approx 0$, so that the spectral 
curvature is fully determined by the quartic term; what is the physical meaning of such a model 
is a mystery to us.

\section{Discussion and summary}

Our main results can be  formulated as follows. 
\begin{itemize}
\item We find that the 20 GeV breaks due to H~LyC are ubiquitous. 
They are statistically significant in the majority of the bright blazars as well as in 
the stacked redshift-corrected spectra. 
\item The 5 GeV breaks due to \heii~LyC remain significant only in two objects. 
\item A more complicated function describing the underlying spectrum can change 
the significance of the break existence. An ad hoc biquadratic function of $\ln E$
gives a slightly better fit that the physically justified lognormal function with the BLR absorption for 3C~454.3, but a much 
worse fit for the stacked blazar spectrum. 
Thus this model does not eliminate the need for the break. 
\item Breaks are not seen in the stacked redshift-corrected BL Lac spectrum or the spectrum of the ``empty'' sky.  
\item The presence of the breaks associated with absorption by UV photons implies 
that at least some fraction of the $\gamma$-rays are produced within   the BLR. 
\end{itemize}
The presence of \heii~LyC absorption in  3C~454.3 is not surprising. This object is exceptional in all its components: 
the $\gamma$-ray emission from the jet reaches luminosities in excess of $2\times 10^{50}$~erg~s$^{-1}$ \citep{Fermi11_3C454}, 
the accretion disk emits  $L_{\rm d}\sim10^{47}$~erg~s$^{-1}$ \citep{Bonnoli11}, the BLR $3\times 10^{45}$~erg~s$^{-1}$ \citep{Pian05}, 
and luminosity in Ly$\alpha$ only is $L_{\rm Ly\alpha}\sim 10^{45}$~erg~s$^{-1}$ \citep{Wills95}. 
Here we can expect that the ionization degree is high and the photon-photon optical depth is substantial.


The second object that shows   \heii~LyC absorption is 4C~+21.35. 
That case is important because this FSRQ has been detected during a flare in the 70--400 GeV range by MAGIC \citep{Magic11_1222} 
and the coexistence of the absorption break and VHE emission is difficult to understand in a single emission zone scenario,
because the multi-hundred GeV photons would have trouble escaping from the BLR.  
The $\gamma$-ray  luminosity of 4C~+21.35 is smaller than in  3C~454.3, $10^{48}$~erg~s$^{-1}$  during flares \citep{Tanaka11}, 
but the accretion disk is almost as luminous with $L_{\rm d}\approx5\times 10^{46}$~erg~s$^{-1}$ \citep{TBG11}. 
On the other hand, the  BLR  luminosity is significantly smaller  $5\times 10^{44}$~erg~s$^{-1}$  
\citep{Wang04,Fan06,Tanaka11} with Ly$\alpha$ producing   $\sim10^{44}$~erg~s$^{-1}$, i.e. ten times less than in 3C~454.3.  
A much  lower BLR luminosity implies that the opacity for 100 GeV photons is smaller, 
because BLR size scales roughly as $R_{\rm BLR}\approx 10^{18} L_{\rm d,47}^{1/2}$~cm \citep{Kaspi07,Bentz09} 
and  the opacity as $\tau_{\rm T}\propto L_{\rm line}/L_{\rm d}^{1/2}$  \citep[see Equation (\ref{eq:taut}) and ][]{PS10}. 
Thus in  4C~+21.35, the optical depth through the BLR is expected to be 5--7 times smaller than in 3C~454.3. 
This might be the reason why VHE emission is detected in 4C~+21.35, but not in a much stronger source 3C~454.3.

For a given object having both the measurement of the line luminosity and the estimation of the BLR size, we 
can obtain an expected value for the  opacity $\tau_{\rm T}$ using Equation~(\ref{eq:taut}), i.e. 
assuming that the BLR emission is isotropic and $\gamma$-rays have to penetrate through the whole BLR.
For 3C~454.3 with $L_{\rm d,47}\sim1$ and $L_{\rm Ly\alpha,45}\sim1$, 
we get the opacity from H LyC/Ly$\alpha$ of $\tau_{\rm H}\sim100$, while for  4C~+21.35 we have $\tau_{\rm H}\approx 15$. 
We see, however, for both objects the observed value is $\tau_{\rm H}\sim$2--5 
(Table~\ref{tab:fits}), much below the expectation. What are the possible solutions for this discrepancy? 
We can propose at least two solutions: 
\begin{itemize}
\item 
The  size of the H LyC/Ly$\alpha$ emission region is larger than that measured from reverberation mapping
using C\,{\sc iv} \citep{Kaspi07}  and H$\beta$ lines \citep{Bentz09}. 
We note that no Ly$\alpha$ variability was detected in any of the quasars analyzed by \citet{Kaspi07}
supporting this picture. 
This would immediately reduce the expected $\gamma$-ray opacity, 
which scales inversely proportionally with the size. 
\item 
Alternatively, 
if the BLR is flat \citep{Shields78,Decarli11}, i.e. elongated alone the accretion disk, 
the $\gamma$-ray opacity is much reduced \citep[e.g.][]{LeiWang14}. 
In this case,  the threshold energy, which depends on the 
maximum interaction angle between the BLR photons and the $\gamma$-rays,  should be a factor of two larger
(i.e. $\sim$40 GeV instead of $\sim$20 GeV) contradicting the data. 
However, if in addition a few BLR clouds are situated along the jet axis outside the $\gamma$-ray emitting region, 
they would produce enough photons to collide head-on with the 
$\gamma$-rays to make a break at 20 GeV. 
\item 
Finally, the  $\gamma$-ray emitting region can be extended and its location can change 
depending on the luminosity, with less GeV absorption during the strong flares  \citep{SP11,Pacciani14}. 
The VHE emission  does not need to be produced in exactly the same place as the GeV emission. 
This can explain the fact that we see both GeV breaks as well as $>$100 GeV emission 
in  4C~+21.35  (but not necessarily at the same time). 
\end{itemize}

How then these models  compares with the results on \heii\ absorption? 
Photoionization models predict  that luminosity in \heii\ LyC/Ly$\alpha$ is typically $\sim$10\% of the hydrogen one 
\citep{TG08,PS10}. 
Thus, taking the same BLR size and geometry 
and using four times larger photon energy gives $\tau_{\rm He}\sim \tau_{\rm H}/40$. 
In that case, the \heii\ absorption would be negligible. 
On the other hand, reverberation mapping shows that \heii\  lines are produced 
closer to the black hole that e.g. H$\beta$ \citep{PW99}, implying a 
higher photon density \textit{inside} the zone of complete He ionization and a larger $\tau_{\rm He}$. 
However,  if the $\gamma$-ray emitting region is slightly outside of that region the opacity is reduced. 
Again in this situation the break energy should  be higher, but 
the quality of the data do not allow us to reject a hypothesis that the \heii\ break 
energy is actually two times larger than the fiducial (rest-frame) 5 GeV value.  
Thus we do not see an obvious contradiction between the fact that 
the $\gamma$-rays are produced outside (or at the edge) of \heii\ LyC emitting region and the presence of the break.

In general, the situation with the H LyC absorption strongly dominating over \heii\ LyC  absorption 
seems more natural than the picture presented by \citet{Abdo10_blazars} and \citet{PS10},  
where a few GeV breaks look more prominent than those at $\sim$20 GeV. 
Now the statement of \citet{PS10} that the jet emission takes place in the inner (higher ionization) regions of BLRs should be modified: 
the gamma-ray emission site lies within the normal \hii\ BLR region. 
This fact does not change the main astrophysical implications of photon-photon absorption of the jet gamma-ray emission, 
this particularly  implies that the jet is already accelerated within a parsec distance from the black hole 
and therefore the Blandford-Znajek mechanism  \citep{BZ77,Komissarov07} is responsible for the jet launching
and the BLR dense photon field can provide conditions for energy dissipation via photon breeding \citep{StP06,SP08}.

\acknowledgments 
This research was supported  by the Academy of Finland grant 268740  and the Magnus Ehrnrooth foundation.
The research made use of public data obtained from the \textit{Fermi} Science Support Center. 



\end{document}